\magnification=\magstep1
\hsize=13cm
\vsize=20cm
\overfullrule 0pt
\baselineskip=13pt plus1pt minus1pt
\lineskip=3.5pt plus1pt minus1pt
\lineskiplimit=3.5pt
\parskip=4pt plus1pt minus4pt

\def\negenspace{\kern-1.1em}



\newcount\secno
\secno=0
\newcount\susecno
\newcount\fmno\def\z{\global\advance\fmno by 1 \the\secno.
                       \the\susecno.\the\fmno}
\def\section#1{\global\advance\secno by 1
                \susecno=0 \fmno=0
                \centerline{\bf \the\secno. #1}\par}
\def\subsection#1{\medbreak\global\advance\susecno by 1
                  \fmno=0
       \noindent{\the\secno.\the\susecno. {\it #1}}\noindent}


\def\sqr#1#2{{\vcenter{\hrule height.#2pt\hbox{\vrule width.#2pt
height#1pt \kern#1pt \vrule width.#2pt}\hrule height.#2pt}}}
\def\square{\mathchoice\sqr64\sqr64\sqr{4.2}3\sqr{3.0}3}


\newcount\refno
\refno=1
\def\y{\the\refno}
\def\myfoot#1{\footnote{$^{(\y)}$}{#1}
                 \advance\refno by 1}


\def\neq{\hbox{$\,$=\kern-6.5pt /$\,$}}





\newcount\secno
\secno=0
\newcount\fmno\def\z{\global\advance\fmno by 1 \the\secno.
                       \the\fmno}
\def\sectio#1{\medbreak\global\advance\secno by 1
                  \fmno=0
       \noindent{\the\secno. {\it #1}}\noindent}


\def\kg{[\![}
\def\gk{]\!]}




\magnification=\magstep1
\hsize 13cm
\vsize 20cm

\hfill {Preprint IMAFF 98/01}
\bigskip\bigskip
\centerline{\bf{OSTROGRADSKI FORMALISM}}
\centerline{\bf{FOR}}
\centerline{\bf{HIGHER-DERIVATIVE SCALAR FIELD THEORIES}}
\vskip 0.3cm 
\centerline{by}
\vskip 0.7cm
\centerline{F.J. de Urries $^{(*)(\dagger)}$}
\vskip 0.2cm
\centerline{and}
\vskip 0.2cm
\centerline{J.Julve $^{(\dagger)}$}
\vskip 0.2cm
\centerline{$^{(*)}$\it Departamento de F\'\i sica, Universidad de Alcal\'a de 
Henares,}
\centerline{\it 28871 Alcal\'a de Henares (Madrid), Spain}
\vskip 0.2cm   
\centerline{$^{(\dagger)}$\it{IMAFF, Consejo Superior de Investigaciones
Cient\'\i ficas,}} 
\centerline{\it{Serrano 113 bis, Madrid 28006, Spain}}

\vskip 0.7cm

\centerline{ABSTRACT}\bigskip 
We carry out the extension of the Ostrogradski method to relativistic field
\break
theories. Higher-derivative Lagrangians reduce to second differential--order 
with one explicit independent field for each degree of freedom. We 
consider a higher-derivative relativistic theory of a scalar field and 
validate a powerful order-reducing covariant procedure by a rigorous phase-space 
analysis. The physical and ghost fields appear explicitly. Our results 
strongly support the formal covariant methods used in higher-derivative gravity.

\bigskip
\centerline{PACS numbers: 11.10.Ef, 11.10.Lm, 04.60} 

\bigskip
\centerline{Keywords: higher-derivative field theories}

\vfill





\eject

\sectio{\bf{Introduction}}\bigskip 
Theories with higher order Lagrangians have an old tradition in physics, and 
Podolski's Generalized Electrodynamics [1] (later visited as a useful testbed  
[2]), effective gravity and tachyons [3] are examples. The interest in higher 
order mechanical systems is alive until today [4].

Theories of gravity with terms of any order in curvatures arise as part of 
the low energy effective theories of the strings [5] and from the dynamics of 
quantum fields in a curved spacetime background [6]. Theories of second order 
(4--derivative theories in the following) have been studied more closely in 
the literature because they are renormalizable [7] in four dimensions. They 
greatly affect the effective potential and phase transitions of scalar fields  
in curved space-time, with a wealth of astrophysical and cosmological 
properties [8]. In particular a procedure based on the Legendre 
transformation was devised [9] to recast them as an equivalent theory of 
second differential order. A suitable diagonalization of the resulting theory 
was found later [10] that yields the explicit independent fields for the
degrees of freedom (DOF) involved, usually including Weyl ghosts. 

In [11] the simplest example of this procedure was given using a model 
of one scalar field with a massless and a massive DOF.
In an appendix, Barth and Christensen [12] gave the splitting of the
higher-derivative (HD) propagator into quadratic ones for the 4th, 6th and 
8th differential-order scalar theories. A scalar 6--derivative theory 
has been considered in [13] as a regularization of the Higgs model, yielding 
a finite theory.

Classical treatises [14] study the Lagrangian and Hamiltonian
theories of systems with a finite number of DOF and higher time-derivatives
of the generalized coordinates. Later work has considered the variational 
problem of those theories with the tools of the Cartan form, k-jets, 
symplectic geometry and Legendre mappings [15].  

However the particular case of relativistic covariant field theories has 
complications of its own which are not covered by those general treatments.
Our presentation highlights the Lorentz covariance and the particle aspect of 
the theory, with emphasis in the structure of the propagators and the 
coupling to other matter sources. We address this issue by using a simplified 
model with scalar fields as in [11] and [12], and our extrapolation of the 
canonical analysis to these continous systems validates the formal 
procedures introduced there. The analysis here presented mostly focuses on 
the free part of the Lagrangian, and self-interactions and interactions with 
other fields are embodied in a source term.

In Section 2 we briefly review the Ostrogradski method and outline our 
extension to the field theories. In Section 3 we study the case of the
4-derivative theory for arbitrary non-degenerate masses, which exemplifies 
the use of the Helmholtz Lagrangian and the crucial diagonalization
of the fields. The 8-derivative case and higher 4$N$-derivative cases are 
considered in Section 4. For even $N$ the 2$N$-derivative cases present some 
peculiarities that deserve the separate discussion of Section 5. Our results    
are summarized in the Conclusions. 

As a general feature, our procedure involves vectors with pure real and 
imaginary components as well as symmetric matrices with equally assorted
elements. Diagonalizing symmetric matrices of this kind is a non-standard 
task which is detailed in an Appendix.

\vfill
\eject

\sectio{\bf The Ostrogradski's method.}
\bigskip

We consider a HD Lagrangian theory for a system described by
configuration variables \quad$q(t)$\quad. By dropping total derivatives, it
can be always brought to a standard form 

$$ L[q,\dot{q},\ddot{q},...,{\buildrel {(m)}\over{q}}]
                                               \quad , \eqno(\z)$$ 

\noindent{depending} on time derivatives of the lowest possible order. 
The variational principle then yields equations of motion which
are of differential order \quad$2m$\quad at most:

$$  {{\partial L}\over{\partial q}}
   -{d\over dt}{{\partial L}\over{\partial{\dot{q}}}}
   +\cdots
   +(-1)^m {d^m \over dt^m}
    {{\partial L}\over{\partial{\buildrel {(m)}\over{q}}}}
   = 0 \quad .                                               \eqno(\z)$$

\noindent{Hamilton's} equations are obtained by defining $m$ generalized
momenta 

$$\eqalign {p_m &\equiv  
      {{\partial L}\over{\partial{\buildrel {(m)}\over{q}}}}  \cr
            p_i &\equiv 
      {{\partial L}\over{\partial{\buildrel {(i)}\over{q}}}}  
      -{d\over dt}p_{i+1}
      \quad\quad (i=1,...,m\! -\! 1)\quad ,  \cr}  \eqno(\z)$$ 

\noindent{and} the \quad$m$\quad independent variables

$$\eqalign {q_1 &\equiv q  \cr
            q_i &\equiv {\buildrel {(i\!-\!1)}\over{q}}  
            \quad\quad (i=2,...,m)\quad .  \cr}  \eqno(\z)$$

\noindent{Then} the Lagrangian may be considered to depend on
the coordinates \quad$q_i$\quad and only on the first time derivative
\quad${\dot q}_m={\buildrel {(m)}\over{q}}$\quad.
A Hamiltonian on the phase space \quad[$q_i,p_i$]\quad may then be found
by working \quad${\dot q}_m$\quad out of the first equation (2.3) as a
function 

$${\bf{\dot q}_m}[q_1,...,q_m;p_m]\quad , \eqno(\z)$$

\noindent{the} remaining velocities \quad${\dot q}_i \;
(i=1,...,m\! -\! 1)$\quad already being expressed in terms of
coordinates, because of (2.4), as

$${\bf {\dot q}_i}=q_{i+1}\quad . \eqno(\z)$$

\noindent{Thus}

$$ H[q_i,p_i]=\sum_{i=1}^{m}\; p_i{\bf{\dot q}_i} - L
               =\sum_{i=1}^{m\!-\!1}\; p_i q_{i+1} + p_m{\bf{\dot q}_m}
                   - L[q_1,...q_m;{\bf{\dot q}_m}]\quad .
                                                      \eqno(\z)$$ 
\noindent{Therefore}

$$\eqalign{\delta H = \sum_{i=1}^{m\!-\!1}&( p_i\delta q_{i+1} + q_{i+1}
\delta p_i)+p_m\delta{\bf {\dot q}_m}+ {\bf {\dot q}_m}\delta p_m  \cr
&-\sum_{i=1}^{m}{{\partial L}\over{\partial q_i}}\delta q_i 
-{{\partial L}\over{\partial {\bf {\dot q}_m}}}\delta{\bf{\dot q}_m}
                                                    \quad,\cr}\eqno(\z)$$ 

\noindent{but} (2.3) can be written as 

$$\eqalign{{\partial L}\over{\partial{\bf{\dot q}_m}} &= p_m  \cr 
      {{\partial L}\over{\partial q_i}}&={\dot p}_i+p_{i-1}
      \quad\quad (i=2,...,m)\quad ,  \cr}  \eqno(\z)$$

\noindent{and} (2.2), because of (2.3), gives

$${{\partial L}\over{\partial q_1}}={{\partial L}\over{\partial q}}
                                           ={\dot p}_1\quad, \eqno(\z)$$

\noindent{so} we get 

$$\delta H= \sum_{i=1}^m(-{\dot p}_i\delta q_i+{\dot q}_i\delta p_i)
                                                       \quad,\eqno(\z)$$

\noindent{and} the canonical equations of motion turn out to be

$$ {\dot q}_i={{\partial H}\over{\partial {p}_i}}\quad ; \quad
    {\dot p}_i=-{{\partial H}\over{\partial {q}_i}}\quad . 
                                                      \eqno(\z)$$

\noindent{Summarizing} we may say that a theory with one
configuration coordinate \quad$q$\quad obeying equations of motion of 
\quad$2m$\quad differential order (stemming from a Lagrangian with quadratic
terms in \quad${\buildrel {(m)}\over{q}}$\quad as its highest derivative
dependence) can be cast as a set of 1st--order canonical equations 
for \quad$2m$\quad phase-space variables \quad$[q_i,p_i]$\quad.

As it is well known, once the differential order has been reduced by 
the Hamiltonian formalism, one may prefer to obtain the same canonical 
equations of motion from a variational principle. Then the canonical 
equations (2.12) are the Euler equations of the so-called Helmholtz 
Lagrangian

$$  L_H[q_i,{\dot q}_i,p_i] = \sum_{i=1}^{m}\; p_i{\dot q}_i 
                                   - H[q_i,p_i] \eqno(\z)$$
 
\noindent{which} depends on the \quad$2m$\quad coordinates \quad$q_i$\quad 
and \quad$p_i$\quad, and 
only on the velocities \quad${\dot q}_i$\quad. This alternative setting will 
be adopted later on.

As far as finite--dimensional mechanical systems are considered, only time 
derivatives are involved. The generalized momenta above have a mechanical 
meaning and the resulting Hamiltonian is the energy of the system up to 
problems of positiveness linked to the occurrence of ghost states.
\bigskip

\noindent{\bf Extension to field theories}

Continuous systems with field coordinates \quad$\phi (t,{\bf x})$\quad 
usually involve space derivatives as well, chiefly if relativistic covariance 
is assumed. We now generalize the previous formalism to this case. A HD field 
Lagrangian density will have the general dependence

$${\cal L}[\phi ,{\phi}_{\mu},...,{\phi}_{{\mu}_1\cdots{\mu}_m}]\quad,
                                                              \eqno(\z)$$ 

\noindent{where}\quad ${\phi}_{{\mu}_1\cdots{\mu}_i}\equiv                           
{\partial}_{\mu_1}\cdots{\partial}_{\mu_i}\phi$\quad, with corresponding
equations of motion

$$  {{\partial{\cal L}}\over{\partial\phi}}
   -{\partial_\mu}{{\partial{\cal L}}\over{\partial{\phi_\mu}}}
   +\cdots
   +(-1)^m{\partial}_{\mu_1}\cdots{\partial}_{\mu_m}
    {{\partial{\cal L}}\over{\partial{\phi}_{{\mu}_1\cdots{\mu}_m}}}
   = 0 \quad .                                               \eqno(\z)$$

\noindent{The} generalized momenta now are

$$ \eqalign{\pi^{{\mu}_1\cdots{\mu}_m}&\equiv 
   {{\partial{\cal L}}\over{\partial{\phi}_{{\mu}_1\cdots{\mu}_m}}}\cr
    \pi^{{\mu}_1\cdots{\mu}_i}&\equiv 
   {{\partial{\cal L}}\over{\partial{\phi}_{{\mu}_1\cdots{\mu}_i}}}
   -\partial_{\mu_{i+1}} \pi^{{\mu}_1\cdots{\mu}_i{\mu_{i+1}}}
    \quad\quad (i=1,...,m\!-\!1)\quad.\cr} \eqno(\z)$$
 
\noindent{Though} they have not a direct mechanical meaning of impulses
they still are suitable to perform a Legendre transformation upon. 

Assuming also that the highest derivative can be worked out of the first 
equation of 
(2.16) as a function \quad 
${\bar\phi}_{{\mu}_1\cdots{\mu}_m}[\phi,\phi_\mu,...,{\phi}_{{\mu}_1\cdots
  \mu_{m-1}};\pi^{{\mu}_1\cdots{\mu}_m}]$\quad, the "Hamiltonian" density 
now is 

$$\eqalign{{\cal H}[\phi,\phi_\mu,...,{\phi}_{{\mu}_1\cdots{\mu}_{m-1}};
                                   \pi^\mu,...,&\pi^{{\mu}_1\cdots{\mu}_m}]
  =\pi^\mu\phi_\mu +\cdots
   + \pi^{{\mu}_1\cdots{\mu}_{m-1}}{\phi}_{{\mu}_1\cdots{\mu}_{m-1}}\cr
   &+\pi^{{\mu}_1\cdots{\mu}_m}{\bar\phi}_{{\mu}_1\cdots{\mu}_m}
   -{\cal L}[\phi,\phi_{\mu},...,{\bar\phi}_{{\mu}_1\cdots{\mu}_m}]\quad.\cr}
                                                            \eqno(\z) $$

\noindent{Then} the canonical equations are

$$\eqalign{\partial_\mu\phi&={{\partial{\cal H}}\over{\partial\pi^\mu}}
                                      \quad,\quad
\partial_\mu\phi_\nu={{\partial {\cal H}}\over{\partial\pi^{\mu\nu}}}
                              \quad,\;...\;,\quad
\partial_\mu{\phi}_{{\mu}_1\cdots{\mu}_{m-1}}=
        {{\partial {\cal H}}\over{\partial\pi^{\mu\mu_1\cdots\mu_{m-1}}}}
                                                          \quad,\cr
\partial_\mu\pi^\mu&=-{{\partial {\cal H}}\over{\partial\phi}} 
                                       \;,\;\;                                                                                                   
\partial_\nu\pi^{\mu\nu}=-{{\partial {\cal H}}\over{\partial\phi_\mu}}
                              \;\;\;,\;...\;,\;\;
\partial_\sigma{\pi}^{{\mu}_1\cdots{\mu}_{m-1}\sigma}=
        -{{\partial {\cal H}}\over{\partial\phi_{\mu_1\cdots\mu_{m-1}}}}
                                                 \;\;.\cr}\eqno(\z)$$

This general setting may be hardly applicable to systems of practical 
interest (generally involving internal symmetries and/or fields belonging to 
less trivial Lorentz representations) if suitable strategies are not adopted 
to refine the method. One crucial observation is that the momenta may be 
defined in more useful and general ways than the plain one introduced in 
(2.16): instead of differentiating with respect to the simple field 
derivatives \quad$\phi_{\mu_1\cdots\mu_i}$\quad one may consider combinations 
of field derivatives of different orders belonging to the same Lorentz and 
internal group representations. For instance, in HD gravity [9], the Ricci
tensor is a most suited combination of second derivatives of the metric 
tensor field. The only condition is that the Lagrangian be regular in the 
highest "velocity" so defined. This will be made clear in the following. 

In fact this general Ostrogradski treatment can be significantly simplified 
for the Lorentz invariant theory of a scalar field, which is the example we 
will consider in this paper. In this case, dropping total derivatives, the 
general form (2.14 ) can be expressed in a more convenient way that singles 
out the free quadratic part, namely

$$ {\cal L} = -{c\over 2} 
              \,\phi\kg1\gk\kg2\gk\cdots\kg N\gk\phi
                                          -j\,\phi\quad,\eqno(\z)$$

\noindent{where}\quad$\kg i\gk \equiv (\square + m^2_i)\;$, our Minkowski 
signature  is \quad$(+,-,-,-)$\quad so that \break 
$\square\equiv\partial^2_t-\triangle\;$, and \quad$c$\quad is a dimensional 
constant. The masses are ordered such that \quad$m_i > m_j$\quad when 
\quad$ i < j$\quad so that the objects \quad$ \langle ij \rangle \equiv 
(m^2_i - m^2_j)$\quad are always positive when \quad$i < j\;$. 

It turns out to be very advantageous to consider only Lorentz 
invariant combinations of  derivatives of the type 
\quad${\square\,}^n\phi$\quad and of the \quad$\phi$\quad field itself with 
suitable dimensional coefficients. Further, it is even more useful to 
consider expressions of the form\quad$\kg i \gk^n\phi\;$. 

\noindent{Thus,} arbitrarily focusing ourselves on \quad$i=1\;$ without loss 
of generality, equation (2.19) may be recast as

$${\cal L}={1\over 2}\sum^N_{n=1}c_n\phi\kg 1\gk^n\phi\,-j\phi\quad,
                                                             \eqno(\z)$$

\noindent{where} the \quad$c_n$\quad are redefined constants.

\noindent{Calling} \quad$m={N\over2}$\quad for even \quad$N\;$, and 
\quad$m={{N+1}\over2}$\quad for 
odd \quad$N\;$, the motion equation now reads

$$\sum^m_{n=1}\kg 1\gk^n{{\partial{\cal L}}\over{\partial(\kg 1\gk^n\phi)}}
     =\sum^N_{n=1}c_n\kg 1\gk^n\phi = j                    \eqno(\z)$$

\noindent{The} Legendre transform can now be performed upon the simpler set 
of {\it generalized momenta}

$$\eqalign{\pi_m &={{\partial{\cal L}}\over{\partial(\kg 1\gk^m\phi)}}  \cr
\pi_{m-1}&={{\partial{\cal L}}\over{\partial(\kg1\gk^{m-1}\phi)}}
                                                        +\kg1\gk\pi_m   \cr
                    \cdots\;\; &\quad\quad\cdots                        \cr
\pi_s &={{\partial{\cal L}}\over{\partial(\kg 1\gk^s\phi)}}+\kg1\gk\pi_{s+1}
                      \quad (s=1,...,m-2)\,. \cr}\eqno(\z)$$

\noindent{The} Hamiltonian will depend on the new phase--space coordinates 
\break $H[\phi_1,...,\phi_m;\pi_1,...,\pi_m]\;$, where 
\quad$\phi_i\equiv\kg1\gk^{i-1}\phi\;$. To this end \quad$\kg1\gk^m\phi$\quad 
has been worked out of the 1st (2.22) for even \quad$N$\quad, or of the 2nd  
(2.22) for odd \quad$N$\quad, in terms of these coordinates.

\noindent{The} dynamics of the system is given by the \quad$2m$\quad 
equations of second order

$$\eqalign{\kg 1\gk \phi_i &= {{\partial H}\over{\partial\pi_i}}\quad\cr
           \kg 1\gk \pi_i &= {{\partial H}\over{\partial\phi_i}}\quad\cr}
                                          (i=1,...,m)\quad.     \eqno(\z)$$         

\noindent{Notice} that, in comparison with (2.12),(2.16) and (2.18), no 
negative sign occurs in both (2.22) and (2.23), because each step now 
involves two derivative orders.

As a final comment, the treatment followed above keeps Lorentz invariance 
explicitely, and this will turn advantageous later on. The price has been 
that neither the \quad$\pi$'s\quad have the meaning of mechanical momenta nor 
\quad$H$\quad has to do with the energy of the system. However they are 
adequate for providing a set of "canonical" equations that correctly describe 
the evolution of the system. Moreover, these equations are Lorentz invariant 
and of 2nd differential order, which will lend itself to an almost direct 
particle interpretation.
\bigskip

One may however choose to work with the genuine Hamiltonian and mechanical 
momenta obtained when the Legendre transformation built-in in the 
Ostrogradski method involves only the true "velocities" 
\quad$\partial_t^n\phi\;$. The price now is loosing the explicit Lorentz 
invariance and  facing more cumbersome calculations, as we will see by an 
example in the 2nd part of the next Section. 
\bigskip
\bigskip

\sectio{\bf N=2 theories.}

These theories allow a particularly simple treatment that will be illustrated 
in the examples \quad$N=2$\quad and \quad$N=4\;$. The equations (2.23) for
\quad$N=2$\quad will now be obtained from a Helmholtz--like Lagrangian of 2nd 
differential order, which is closer to a direct particle interpretation.

Consider the \quad$N=2$\quad Lagrangian

$$ {\cal L}^{4} = -{1 \over 2}{1 \over M}\,\phi\kg 1\gk\kg 2\gk\phi
                                          -j\,\phi\quad.\eqno(\z)$$

\noindent{with} non-degenerate masses \quad$m_1>m_2\;$. Taking the 
dimensional constant \break $M=(m_1^2 - m_2^2)\equiv \langle 12 \rangle 
>0\;$, equation (3.1) yields the propagator

$$-{{\langle 12 \rangle}\over{\kg 1\gk\kg 2\gk}} 
      = {1 \over{\kg 1\gk}}-{1 \over{\kg 2\gk}}
                                        \quad, \eqno(\z)$$

\noindent{We} thus see that the pole at \quad$m_2$\quad then corresponds to a 
physical particle and the one at \quad$m_1$\quad to a negative norm 
"poltergeist". The 2nd order Lagrangian we are seeking should describe two 
fields with precisely the particle propagators occurring in the r.h.s. of 
(3.2).

The Lagrangian (3.1) can be brought to the form (2.20), namely

 $$ \eqalign{{\cal L}^{4}[\phi,\kg 1\gk\phi]
                    &=-{1 \over 2}{1 \over{\langle 12 \rangle}}
                                      \bigl[\phi\kg 1\gk^2\phi 
             - \langle 12 \rangle \phi\kg 1\gk\phi\bigl]
                                    -j\,\phi       \cr
                    &=-{1 \over 2}{1 \over{\langle 12 \rangle}}
                                      \bigl[(\kg 1\gk\phi)^2 
             - \langle 12 \rangle \phi(\kg 1\gk\phi)\bigl]
                                    -j\,\phi\quad ,\cr}\eqno(\z)$$

\noindent{where} the relationship \quad$\kg 2\gk=\kg 1\gk-\langle 12 \rangle$ 
\quad has been used. 

We define one momentum

$$\pi={{\partial{\cal L}}\over{\partial(\kg 1\gk\phi)}}\eqno(\z)$$

\noindent{from} which \quad$\kg 1\gk\phi$\quad is readily worked out, 
obtaining

$${\cal H}^4[\phi,\pi]=-{1\over 2}\langle 12 \rangle(-\pi+{1\over 2}\phi)^2
                                                    +j\,\phi\eqno(\z)$$
\noindent{and} the Helmholtz-like Lagrangian is

$${\cal L}^4_H[\phi,\kg 1\gk\phi,\pi]=\pi\kg 1\gk\phi- {\cal H}[\phi,\pi]
                                                        \quad.\eqno(\z)$$

\noindent{It} contains mixed terms \quad$\pi\,\phi$\quad that obscure the 
particle contents. The diagonalization is achieved by new fields 
\quad$\phi_1\;$ , $\;\phi_2$
 
$$\eqalign{\phi&= \phi_1 + \phi_2 \cr 
            \pi&= {1\over2}(\phi_1 - \phi_2)\cr} \eqno(\z)$$

\eject
\noindent{to} yield

$${\cal L}^{2}={1\over2}\phi_1\kg 1\gk \phi_1
                -{1\over2}\phi_2\kg 2\gk \phi_2
                -j(\phi_1+\phi_2)\quad ,\eqno(\z)$$

\noindent{where} the particle propagators in the r.h.s. of (3.2) are 
apparent. This result is physically meaningful: where we had a single field 
\quad$\phi\;$, coupled to a source \quad$j\;$, propagating with the quartic 
propagator in the l.h.s. of (3.2) as implied by the HD Lagrangian (3.1), we 
now have two fields \quad$\phi_1\;$ , $\;\phi_2$\quad describing particles 
with quadratic propagators, and the source couples to the sum 
\quad$\phi_1+\phi_2\;$.
\bigskip

A deeper insight of the phase-space structure of the theory can be achieved 
by the plain use of the Ostrogradski method, eventually confirming the final
form (3.8). In order to explicitely show the velocities, we write (3.1) in 
the form of the Lagrangian density

$${\cal L}^{4}=-{1\over 2}{1\over{\langle 12\rangle}}
      \{(\partial^2_t\phi)^2-(\partial_t\phi)S(\partial_t\phi)
                             + \phi P\phi\} -j\,\phi \eqno(\z)$$ 

\noindent{where} \quad$S\equiv M^2_1+M^2_2\;,\;P\equiv M^2_1M^2_2$\quad and
 \quad$M^2_i\equiv m^2_i-\triangle$\quad are operators containing the space 
derivatives.

The Ostrogradski formalism yields the Hamiltonian density

$$ {\cal H}^{4}[\phi,\dot\phi;\pi_1,\pi_2]= -{1\over 2}\langle 
12\rangle\pi^2_2
   +\pi_1\dot\phi-{1\over 2}{1\over{\langle 12\rangle}}\dot\phi S\dot\phi
   +{1\over 2}{1\over{\langle 12\rangle}}\phi P\phi +j\,\phi    \eqno(\z)$$

\noindent{that} depends on the phase-space coordinates 
\quad$\phi\;,\;\dot\phi\;,\;\pi_1\;,\;\pi_2$\quad and on their space 
derivatives. The highest-order "velocity" \quad$\partial^2_t\phi$\quad has 
been worked out of the momenta

$$\eqalign{\pi_2&\equiv
       {{\partial {\cal L}^4}\over{\partial(\partial^2_t\phi)}}
           =-{1\over{\langle 12\rangle}}\partial^2_t\phi\quad ,\cr
           \pi_1&\equiv
       {{\partial {\cal L}^4}\over{\partial(\partial_t\phi)}}
                                    -\partial_t\pi_2\quad . \cr}\eqno(\z)$$

\noindent{The} canonical equations may be derived from the Helmholtz 
Lagrangian

$$\eqalign{ 
   {\cal L}^4_H[\phi,\dot\phi;\pi_1\pi_2;\partial_t\phi,\partial_t\dot\phi]         
   =\pi_2\partial_t\dot\phi+\pi_1\partial_t\phi
    & + {1\over 2}\langle 12\rangle\pi^2_2
    -\pi_1\dot\phi+{1\over 2}{1\over{\langle 12\rangle}}\dot\phi S\dot\phi\cr
    &-{1\over 2}{1\over{\langle 12\rangle}}\phi P\phi -j\,\phi\quad.\cr}                                                                  
\eqno(\z)$$

\noindent{This} is a Lagrangian density of 1st order in time derivatives,
and we express it in matrix form for later convenience:

$$ {\cal L}^4_H = {1\over 2}\Phi^T\mu\,\Sigma\,\partial_t\Phi
                 +{1\over 2}\Phi^T{\cal M}_4\,\Phi-J^T\,\Phi\quad,\eqno(\z)$$

\noindent{where} \quad$\mu$\quad is an arbitrary mass parameter and

$$\eqalign{\Phi&\equiv
      \left(\matrix{\pi_2\cr\mu^{-1}\dot\phi\cr\mu^{-1}\pi_1\cr\phi}\right)
                                               \quad,\quad
 \Sigma\equiv
       \left(\matrix{0&1&0&0\cr-1&0&0&0\cr0&0&0&1\cr0&0&-1&0\cr}\right)
                                                                  \quad,\cr
 {\cal M}_4&\equiv
      \left(\matrix{\langle 12\rangle&0&0&0\cr
                         0&{{\mu^2S}\over{\langle 12\rangle}}&-\mu^2&0\cr
                         0&-\mu^2&0&0\cr
                         0&0&0&-{P\over{\langle 12\rangle}}\cr}\right)
                                              \quad,\quad
     J\equiv\left(\matrix{0\cr0\cr0\cr j}\right)\quad,\cr}\eqno(\z)$$

\noindent{with} mass dimensions \quad$[\Phi]=1\;,\; [{\cal M}_4]=2$\quad and 
\quad$[J]=3\;$.

In order to relate (3.13) to (3.8), we have to convert the latter into a 1st 
order theory as well. This is readily done by expressing the velocities
\quad$\partial_t\phi_1$\quad and \quad$\partial_t\phi_2$\quad in terms of the 
momenta

$$\eqalign{{\tilde\pi}_1&\equiv
       {{\partial {\cal L}^2}\over{\partial(\partial_t\phi_1)}}
           =-\partial_t\phi_1\quad ,\cr
           {\tilde
\pi}_2&\equiv
       {{\partial {\cal L}^2}\over{\partial(\partial_t\phi_2)}}
           =\partial_t\phi_2\quad , \cr}\eqno(\z)$$

\noindent{so} that

$${\cal H}^2[\phi_1,\phi_2,{\tilde\pi}_1,{\tilde\pi}_2]
   = -{1\over2}{\tilde\pi}_1^2+{1\over2}{\tilde\pi}_2^2
      -{1\over2}\phi_1M^2_1\phi_1+{1\over2}\phi_2M^2_2\phi_2
                            +j\,(\phi_1+\phi_2)\quad.\eqno(\z)$$

The Helmholtz Lagrangian that yields the canonical equations now is

$$ {\cal L}^2_H = {1\over 2}\Theta^T\mu\,\Sigma\,\partial_t\Theta
     +{1\over 2}\Theta^T{\cal M}_2\,\Theta-J^T{\cal Z}\,\Theta\eqno(\z)$$
 
\noindent{where}
$$\Theta\equiv
      \left(\matrix{\mu^{-1}{\tilde\pi}_1\cr\phi_1\cr\mu^{-1}{\tilde\pi}_2                                                           
\cr\phi_2}\right)
                                               \quad,\quad
{\cal M}_2\equiv
      \left(\matrix{\mu^2&0&0&0\cr
                     0&M^2_1&0&0\cr
                     0&0&-\mu^2&0\cr
                     0&0&0&-M^2_2\cr}\right)
                                           \quad,\quad \eqno(\z)$$

\noindent{with} mass dimensions \quad$[\Theta]=1$\quad and
\quad$[{\cal M}_2]=2\;$, and \quad${\cal Z}$\quad is any matrix with the 
fourth row equal to \quad$(0,1,0,1)\;$.  

The field redefinition analogous to the diagonalizing equations 
(3.7) now is a \quad$4\times4$\quad mixing of fields given by

$$ \Phi= {\cal X}\,\Theta   \eqno(\z)$$

\noindent{where} the invertible matrix

$${\cal X}\equiv \left(\matrix{0&-{{M^2_1}\over{\langle 12\rangle}}&0&
                                      -{{M^2_2}\over{\langle 12\rangle}}\cr
                     -1&0&1&0\cr
                     -{{M^2_2}\over{\langle 12\rangle}}&0&
                      {{M^2_1}\over{\langle 12\rangle}}&0\cr
                      0&1&0&1\cr}\right)                     \eqno(\z)$$

\noindent{verifies}

            $${\cal X}^T\Sigma\,{\cal X}=\,\Sigma    \eqno(\z)$$

            $${\cal X}^T{\cal M}_4{\cal X}=\,{\cal M}_2 \eqno(\z)$$

\noindent{so} we can identify \quad${\cal Z}={\cal X}\;$.

We thus see that (3.19) translates (3.13) into (3.17), and therefore the 
Lagrangians (3.9) and (3.8) are again seen to be equivalent. The derivation 
of the matrix \quad${\cal X}$\quad is cumbersome but contains interesting 
details that worth the Appendix. Notice that the components of 
\quad$\Phi$\quad are expressed by (3.19) in terms of the components of 
\quad$\Theta$\quad {\it and} of their space derivatives. This is not 
surprising as long as \quad$\pi_1\;$, given by (3.11), contains space 
derivatives of \quad$\phi$\quad as well.

Though the plain non-covariant Ostrogradski method we have just implemented 
eventually shows up the Lorentz invariance, the readiness of the explicitely 
covariant procedure  formerly introduced in this Section is apparent. The 
non-covariant approach using  the canonical Hamiltonian and mechanical 
momenta is rigourous and validates the former, but involves more bulky 
diagonalizing matrices with elements that contain space derivatives.     
\bigskip

\sectio{\bf N=4  and higher even N theories}

We treat the \quad$N=4$\quad Theory with the far more practical Lorentz 
invariant method of the previous Section. Otherwise one would have to face 
the diagonalization of \quad$8\times 8$\quad matrices analogous to 
\quad${\hat{\cal M}}_2$\quad and \quad${\hat{\cal M}}_4$\quad in Appendix A. 
Our Lagrangian now is

$${\cal L}^8=-{1\over 2}{{\mu^6}\over M}
            \phi\kg 1\gk\kg 2\gk \kg 3\gk\kg 4\gk\phi-j\,\phi \eqno(\z)$$

\noindent{where} the mass dimensions 
\quad$[\mu]=[\phi]=1\quad,\quad[M]=12$\quad and \quad$[j]=3$\quad
are such that \quad$[{\cal L}^8]=4\;$. Taking \quad$M=\langle12\rangle 
\langle13\rangle
\langle14\rangle \langle23\rangle \langle24\rangle \langle34\rangle\;$, 
equation (4.1) treats the masses \quad$m_i\,(i=1,...,4)$\quad on an equal 
footing, which is apparent in the propagator

$$-{{\mu^{-6}M}\over{\kg 1\gk\kg 2\gk \kg 3\gk\kg 4\gk}}
     = {{\langle1\rangle}\over{\kg1\gk}}-{{\langle2\rangle}\over{\kg2\gk}}
      +{{\langle3\rangle}\over{\kg3\gk}}-{{\langle4\rangle}\over{\kg4\gk}}
                                                              \eqno(\z)$$

\noindent{where} \quad${\langle i\rangle}\equiv\mu^{-6}M\prod\limits_{j\neq 
i}
 {1\over{\langle ij\rangle}}$\quad (remind the ordering convention 
\quad$i<j\;$) with
mass dimensions \quad$[\langle i\rangle]=0\;$.

As for (3.2), the propagator expansion (4.2) suggests that the    
lower-derivative equivalent theory should now be 

$$\eqalign{{\cal L}^{2}={1\over2}{1\over{\langle1\rangle}}\phi_1\kg 1\gk 
\phi_1
                -{1\over2}{1\over{\langle2\rangle}}\phi_2\kg 2\gk \phi_2
                &+{1\over2}{1\over{\langle3\rangle}}\phi_3\kg 3\gk \phi_3
                -{1\over2}{1\over{\langle4\rangle}}\phi_4\kg 4\gk \phi_4\cr
                &-j(\phi_1+\phi_2+\phi_3+\phi_4)\quad .\cr}\eqno(\z)$$

\noindent{We} derive this Lagrangian from (4.1) in the following. In matrix 
form, (4.3) reads

$${\cal L}^{2}={1\over2}\tau^T\kg 1\gk I\tau+{1\over2}\tau^T{\cal M}_2\tau
                   -J^T\,F\tau \quad, \eqno(\z)$$

\noindent{where}

$$\tau\equiv\left(\matrix{\langle1\rangle^{-{1\over 2}}\phi_1\cr
                        -i\langle2\rangle^{-{1\over 2}}\phi_2\cr
                          \langle3\rangle^{-{1\over 2}}\phi_3\cr
                        -i\langle4\rangle^{-{1\over 2}}\phi_4\cr}\right)
                                             \; ,\;
J\equiv\left(\matrix{0\cr0\cr0\cr j}\right)
                                             \; ,\;
{\cal M}_2\equiv\left(\matrix{0&0&0&0\cr
                              0&-\langle12\rangle&0&0\cr 
                              0&0&-\langle13\rangle&0\cr
                              0&0&0&-\langle14\rangle\cr}\right)
                                              \quad ,\eqno(\z)$$

\noindent{$I$}\quad is the \quad$4\times 4$\quad identity, and \quad$F$\quad 
is any matrix with the fourth row equal to \break$(\langle1\rangle^{1\over 
2},i\langle2\rangle^{1\over 2},
     \langle3\rangle^{1\over 2},i\langle4\rangle^{1\over 2})\;.$

By dropping total derivatives we express (4.1) in a standard form involving 
derivatives of the lowest possible order, namely

$$\eqalign{{\cal L}^8[\phi,\kg 1\gk\phi,\kg 1\gk^2\phi]
        =-{1\over 2}{{\mu^6}\over M}
         \{(\kg 1\gk^2\phi)^2&-S(\kg 1\gk\phi)(\kg 1\gk^2\phi)
            +p(\kg 1\gk\phi)^2\cr &-P\phi(\kg 1\gk\phi)\} - j\,\phi\quad,\cr} 
\eqno(\z)$$

\noindent{where}\quad$S\equiv\langle12\rangle+\langle13\rangle+\langle14
\rangle \quad,\quad
   p\equiv\langle12\rangle\langle13\rangle+\langle12\rangle\langle14\rangle                                                                                   
+\langle13\rangle\langle14\rangle\;$, and\break  
$P\equiv\langle12\rangle\langle13\rangle\langle14\rangle\;$.

Ostrogradski-like momenta are defined as follows

$$\eqalign{\pi_2&={{\partial{\cal L}^8}\over{\partial(\kg 1\gk^2\phi)}}
          =-{{\mu^6}\over M}(\kg 1\gk^2\phi)+{{\mu^6S}\over 2M}\kg1\gk\phi\cr
           \pi_1&={{\partial{\cal L}^8}\over{\partial(\kg 1\gk\phi)}}
                   +\kg 1\gk\pi_2\quad .\cr} \eqno(\z)$$

From the 1st of (4.7) the highest derivative is worked out, namely

$$ {\bf \kg 1\gk^2\phi}[\pi_2\,,\kg 1\gk\phi]
                 =-{M\over{\mu^6}}\pi_2+{S\over 2}(\kg 
1\gk\phi)\quad.\eqno(\z)$$

The "Hamiltonian" functional is

$${\cal H}^8[\psi_1,\psi_2,\pi_1,\pi_2]
                =\pi_2{\bf \kg 1\gk^2\phi}+\pi_1\psi_2
                 -{\cal L}^8[\psi_1,\psi_2,{\bf \kg 
1\gk^2\phi}]\quad,\eqno(\z)$$ 

\noindent{where} \quad$\psi_1\equiv \phi$\quad and \quad$\psi_2\equiv
\kg 1\gk\phi\;$. Its canonical equations can be derived from the Lagrangian

$${\cal L}^8_H={1\over 2}\Phi^T\kg 1\gk{\cal K}\Phi
                +{1\over 2}\Phi^T{\cal M}_8\Phi-J^T\Phi\quad, \eqno(\z)$$

\noindent{where}\quad$J$\quad is the same as in (4.5),

$$\eqalign{\Phi&\equiv\left(\matrix{\mu^2\pi_2\cr
                          \mu^{-2}\psi_2\cr
                          \pi_1\cr
                          \psi_1\cr}\right)
                                            \quad ,\quad
{\cal K}\equiv\left(\matrix{0&1&0&0\cr
                            1&0&0&0\cr
                            0&0&0&1\cr
                            0&0&1&0\cr}\right)
                                            \quad {\rm and} \cr
 {\cal M}_8&\equiv\left(\matrix{\mu^{-10}M&-{S\over 2}&0&0\cr
     -{S\over 2}&-{\mu^{-10}\over M}(p-{{S^2}\over 4})&-\mu^2&
                         {{\mu^2}\over{2\langle 1\rangle}}\cr
                                              0&-\mu^2&0&0\cr
                   0&{{\mu^2}\over{2\langle 1\rangle}}&0&0\cr}\right)
                                                 \quad .\cr}\eqno(\z)$$ 

{\it Prior} to its diagonalization we write (4.10) in the form

$${\cal L}^8_H={1\over2}\Omega^T\kg 1\gk I\Omega
              +{1\over2}\Omega^T{\hat{\cal M}}_8\Omega
                  -J^T{\cal D}^T\Omega\quad,  \eqno(\z)$$

\noindent{where} \quad$\Omega\equiv({\cal D}^T)^{-1}\Phi\;$, with

$$ {\cal D}\equiv {1\over{\sqrt{2}}}\left(\matrix{1&1&0&0\cr
                                                 -i&i&0&0\cr
                                                  0&0&1&1\cr
                                                  0&0&-i&i\cr}\right)
                                                              \eqno(\z)$$
\noindent{and}

$$ {\hat{\cal M}}_8\equiv{\cal D}{\cal M}_8{\cal D}^{-1}=
   {1\over 2}\left(\matrix{M_--S&-iM_+&-\mu^21_-&i\mu^21_+\cr
                           -iM_+&-(M_-+S)&-i\mu^21_-&-\mu^21_+\cr
                             -\mu^21_-&-i\mu^21_-&0&0\cr
                             i\mu^21_+ &-\mu^21_+&0&0\cr}\right)
                                                              \eqno(\z)$$

\noindent{with} \quad$M_{\pm}\equiv{M\over{\mu^{10}}}\pm{{\mu^{10}}\over M}
                   (p-{{S^2}\over 4})$ \quad and 
               \quad$1_{\pm}\equiv 1\pm {1\over{2\langle 1\rangle}}\;$.

Now the task is to establish the equivalence of (4.12) and (4.4). One may 
first check that the eigenvalues \quad$\lambda_i\,(i=1,...,4)$\quad of 
\quad${\hat{\cal M}}_8$\quad
are the diagonal elements of \quad${\cal M}_2$\quad in (4.5). The orthogonal 
matrix \quad$T$\quad
that diagonalizes \quad${\hat{\cal M}}_8$\quad is obtained by working out its 
orthonormal eigenvectors \quad$\mid\!\lambda_i\rangle$\quad with the suitable 
sign, and
arranging them as the columns. These are

$$\eqalign{
\mid\lambda_1\rangle&={{\langle 1\rangle^{1\over 2}}\over{\sqrt 2}}
                         \left(\matrix{0\cr0\cr1_+\cr-i1_-\cr}\right)
                                               \quad ,\cr
\mid\lambda_j\rangle&=
{{i^{(1-\delta_{3j})}\langle j\rangle^{1\over2}}\over
                      {\sqrt 2[-{2\over{\mu^{10}}}M+2\langle 1j\rangle-S]}}
\left(\matrix{{2\over{\mu^{2}}}[-{{\mu^4}\over{\langle 1\rangle}}
          +\langle 1j\rangle(2\langle 1j\rangle-S-M_-)]\cr
          i{2\over{\mu^{2}}}[-{{\mu^4}\over{\langle 1\rangle}}
                                +\langle 1j\rangle M_+]\cr
                 1_-[-2\mu^{-10}M+2\langle 1j\rangle-S]\cr
              -i 1_+[-2\mu^{-10}M+2\langle 1j\rangle-S]\cr}\right)
                                                   \quad ,\cr}\eqno(\z)$$

\noindent{where} \quad$j=2,3,4\;.$ If \quad$I$\quad is the identity matrix, 
we therefore have

$$ T^T\,I\,T=\,I \quad ,\quad 
                    T^T\,{\hat{\cal M}}_8\,T={\cal M}_2\quad ,\eqno(\z)$$                                        

\noindent{and} the fourth row of \quad${\cal D}^TT$\quad can be seen to be 
\quad$(\langle1\rangle^{1\over 2},i\langle2\rangle^{1\over 2},
  \langle3\rangle^{1\over 2},i\langle4\rangle^{1\over 2}\,)\;$, i.e. it has 
the required form for \quad$F\;$. Then, by taking \quad$\Omega=T\tau\;$, 
(4.12) is identical to (4.4).
\bigskip

The general case for even \quad$N\geq6$\quad in the covariant treatment would 
involve
\quad${N\over 2}$\quad Ostrogradski--like momenta and the diagonalization of 
a \quad$N\times N$\quad mass matrix. The non--covariant Ostrogradski method 
introduced in Section 3, which reduces the theory to a 1st differential--
order form,
would now involve \quad$2N\times 2N$\quad matrices. In both treatments the 
procedure would follow analogous paths, albeit with the occurrence of 
intractable eigenvector and diagonalization problems.  
\eject

\sectio{\bf N=3 and higher odd N theories.}

For \quad$N=3\;$, the HD Lagrangian

$$ {\cal L}^6=-{1\over 2}{{\mu^2}\over M}\phi\kg1\gk\kg2\gk\kg3\gk\phi
                                              -j\,\phi  \quad , \eqno(\z)$$  

\noindent{where} 
\quad$M\equiv\langle12\rangle\langle13\rangle\langle23\rangle$\quad
and \quad$[{\cal L}^6]=4\;$, yields the propagator

$$-{{\mu^{-2}M}\over{\kg1\gk\kg2\gk\kg3\gk}}
         =-{{\mu^{-2}\langle23\rangle}\over{\kg1\gk}}
          +{{\mu^{-2}\langle13\rangle}\over{\kg2\gk}}
          -{{\mu^{-2}\langle12\rangle}\over{\kg3\gk}} \quad. \eqno(\z)$$

\noindent{Then}, the expected equivalent 2nd--order theory is 

$${\cal L}^2=-{1\over 2}{{\mu^2}\over{\langle23\rangle}}\phi_1\kg1\gk\phi_1
             +{1\over 2}{{\mu^2}\over{\langle13\rangle}}\phi_2\kg2\gk\phi_2
             -{1\over 2}{{\mu^2}\over{\langle12\rangle}}\phi_3\kg3\gk\phi_3
                                 -j(\phi_1+\phi_2+\phi_3)\,. \eqno(\z)$$

Already for \quad$N=3\;$, the non--covariant Ostrogradski method becomes 
exceedingly cumbersome. In fact, it reduces both (5.1) and (5.3) to
1st differential order in time. Proving the equivalence of those theories 
then involves the diagonalization of \quad$6\times 6$\quad matrices (the 
counterpart
of \quad${\hat{\cal M}}_4$\quad and \quad${\hat{\cal M}}_2$\quad in (A.4)), 
although with a reasonable amount of work it can still be checked that both 
mass matrices have the same eigenvalues, namely \quad$\pm\mu M_1$\quad, 
\quad$\pm\mu M_2$\quad and \quad$\pm\mu M_3\;$.
Finding the eigenvectors and building up the compound diagonalizing 
transformation does not worth the effort.

For the odd \quad$N$\quad theories, the covariant method exhibits an 
interesting feature. Without loss of generality we again single out the 
Klein-Gordon operator \quad$\kg 1\gk$\quad and write (5.1) as

$${\cal L}^6[\phi,\kg1\gk\phi,\kg1\gk^2\phi]
            =-{1\over 2}{{\mu^2}\over M}
              \{(\gk1\gk\phi)(\gk1\gk^2\phi)-S(\gk1\gk\phi)^2
                +P\phi(\gk1\gk\phi)\}-j\,\phi\quad, \eqno(\z)$$

\noindent{where} now \quad$S\equiv\langle12\rangle+\langle13\rangle$\quad and 
\quad$P\equiv\langle12\rangle\langle13\rangle\;$.

The momenta are

$$\eqalign{\pi_2&={{\partial{\cal L}^6}\over{\partial(\kg1\gk^2\phi)}}
                 =-{1\over 2}{{\mu^2}\over M}\kg1\gk\phi  \cr
           \pi_1&={{\partial{\cal L}^6}\over{\partial(\kg1\gk\phi)}}                                                      
+\kg1\gk\;\pi_2 =-{{\mu^2}\over M}\kg1\gk^2\phi+{{\mu^2}\over M}S\kg1\gk\phi 
-{1\over 2}{{\mu^2}\over M}P\phi\quad .\cr}                                                                                                                               
\eqno(\z)$$
\eject

\noindent{Unlike} in (4.7), the highest derivative now is worked out of  
\quad$\pi_1$\quad (instead of \quad$\pi_2\;$), namely

$$\kg1\gk^2\phi[\phi,\kg1\gk\phi,\pi_1]=-{M\over{\mu^2}}\pi_1
                            +S\kg1\gk\phi-{1\over 2}P\phi\quad , \eqno(\z)$$   

\noindent{and}, in terms of the coordinates                                      
\quad$\pi_1\;,\pi_2\;,\psi_1\equiv\phi$\quad and
               \quad$\psi_2\equiv\kg1\gk\phi\;$, the "Hamiltonian" reads  

$${\cal H}^6[\psi_1,\psi_2,\pi_1,\pi_2]=\pi_2\kg1\gk^2\phi+\pi_1\psi_2
                 -{\cal L}^6[\psi_1,\psi_2,\kg1\gk^2\phi]\quad.\eqno(\z)$$

\noindent{The} Helmholtz Lagrangian is

$$\eqalign{{\cal 
L}^6_H[\psi_1,\psi_2,\pi_1,\pi_2]=\pi_2\kg1\gk\psi_2+\pi_1\kg1\gk\psi_1
          &+{M\over{\mu^2}}\pi_1\pi_2-S\pi_2\psi_2+{1\over 2}P\pi_2\psi_1\cr
          &-{1\over 2}\pi_1\psi_2
              -{{\mu^2}\over 4M}P\psi_1\psi_2-j\psi_1\quad.\cr}\eqno(\z)$$

The distinctive feature of the odd \quad$N$\quad cases is that the 1st of 
(5.5), namely 
\quad$\pi_2=-{1\over 2}{{\mu^2}\over M}\psi_2\;$, is a constraint that 
guarantees the relationship \quad$\kg1\gk\psi_1=\psi_2\;$, so one just has 
\quad$N$\quad degrees of freedom.  For even \quad$N$\quad it arises directly 
as an equation of motion. Moreover, unlike the Dirac Lagrangian for spin-
${1\over2}$ fields or the constraints introduced by means of  multipliers, 
the constraint above can be freely imposed on the Lagrangian since it does 
not eliminate the dependence on the remaining variables \quad$\psi_1$\quad 
and \quad$\pi_1\;$. Thus, (5.8) can be expressed in terms of only the three 
fields \quad$\psi_1\;$, $\;\pi_1$\quad and \quad$\pi_2\;$:    

$${\cal L}^6_H[\psi_1,\pi_1,\pi_2]={1\over 2}\Phi^T\kg 1\gk{\cal K}'\Phi
                +{1\over 2}\Phi^T{\cal M}_3\Phi-J^T\Phi\quad ,\eqno(\z)$$

\noindent{where}

$$\eqalign{\Phi&\equiv\left(\matrix{\mu^2\pi_2\cr
                                         \pi_1\cr
                                          \phi\cr}\right) \quad ,\quad
             J\equiv\left(\matrix{0\cr 0\cr j\cr}\right)     \quad ,\cr
{\cal K}'&\equiv\left(\matrix{-4{M\over{\mu^6}}&0&0\cr
                                              0&0&1\cr
                                              0&1&0\cr}\right)\quad , \cr
{\cal M}_3&\equiv
\left(\matrix{4{MS\over{\mu^6}}&2{M\over{\mu^4}}&{P\over{\mu^2}}\cr
                                            2{M\over{\mu^4}}&0&0\cr
                                             {P\over{\mu^2}}&0&0\cr}\right) 
                                                              \quad .\cr}
                                                                \eqno(\z)$$

\noindent{The} Lagrangian (5.9) is expected to be equivalent to (5.3), which 
in
matrix form reads

$${\cal L}^2=-{1\over 2}\tau^T\kg 1\gk I\tau 
             +{1\over 2}\tau^T{\cal M}'_2\tau - J^TG\tau\quad , \eqno(\z)$$

\noindent{where} \quad$I$\quad is the \quad$3\times 3$\quad identity matrix,

$$\tau\equiv\left(\matrix{\mu\langle 23\rangle^{-{1\over 2}}\phi_1\cr 
                         i\mu\langle 13\rangle^{-{1\over 2}}\phi_2\cr
                          \mu\langle 12\rangle^{-{1\over 2}}\phi_3\cr}
                                                  \right)\quad , \quad
{\cal M}'_2\equiv\left(\matrix{0&0&0\cr
                             0&\langle 12\rangle&0\cr
                             0&0&\langle 13\rangle\cr}\right)\quad,
                                                          \eqno(\z)$$

\noindent{and} $\;G\;$ is any matrix with the third row given by
                $\;(\mu^{-1}\langle 23\rangle^{1\over 2},
                -i\mu^{-1}\langle 13\rangle^{1\over 2},
                  \mu^{-1}\langle 12\rangle^{1\over 2})\;$.

The transformation of (5.9) into (5.11) is performed by the field 
redefinition

$$\Phi={\cal D}'T\tau  \quad ,\eqno(\z)$$

\noindent{where}

$${\cal D}'\equiv{1\over{\sqrt 2}}
           \left(\matrix{{{\mu^3}\over{\sqrt{2M}}}&0&0\cr
                                               0&-i&-1\cr
                                               0&-i&1 \cr}\right)\quad ,
                                                                \eqno(\z)$$

\noindent{and} \quad$T$\quad is an orthogonal matrix built up with the 
eigenvectors of \quad${\cal D}'^T{\cal M}_3{\cal D}'\;$, namely

$$T={{\mu}\over{2\sqrt 2}}\langle 23\rangle^{-{1\over 2}}
  \left(\matrix{0&i{{2\sqrt 2}\over{\mu}}\langle 12\rangle^{1\over 2}
                 &-{{2\sqrt 2}\over{\mu}}\langle 13\rangle^{1\over 2}\cr
                  -i{{P_-}\over P}
                 &{{P_+}\over{\sqrt M}}\langle 12\rangle^{-{1\over 2}}
                 &i{{P_+}\over{\sqrt M}}\langle 13\rangle^{-{1\over 2}}\cr
                  {{P_+}\over P}
                 &i{{P_-}\over{\sqrt M}}\langle 12\rangle^{-{1\over 2}}
                 &-{{P_-}\over{\sqrt M}}\langle 13\rangle^{-{1\over 2}}\cr}
                                            \right)\quad ,\eqno(\z)$$

\noindent{with} \quad$P_{\pm}\equiv P\pm\mu^{-2}2M\;$. 

Then \quad${\cal D}'^T{\cal K}'{\cal D}'=-I$\quad and
\quad$T^T{\cal D}'^T{\cal M}_3{\cal D}'T={\cal M}'_2\;$. One may also check 
that \quad${\cal D}'T$\quad has the same third row required for \quad$G\;$.

\bigskip

The covariant treatment of the general odd \quad$N\geq 5$\quad case proceeds 
along hte same lines. Initially \quad$(N+1)/2$\quad Ostrogradski coordinates 
plus the corresponding momenta occur. Again the definition of the highest 
momentum yields a constraint with the same meaning as above, while the 
highest field derivative is worked out of the next momentum definition. Then 
one faces the diagonalization of a Helmholtz Lagrangian depending on just 
\quad$N$\quad fields. 

Already in the \quad$N=3$\quad case one might have chosen not to implement 
the constraint on the Lagrangian (5.8) and let it to arise in the equations 
of motion. These equations are the canonical ones for the Hamiltonian (5.7) 
and involve an even number of variables, as required by phase space. Thus one 
keeps the dependence of the Lagrangian (5.8) on the four fields
\quad$\psi_1\;,\;\psi_2\;,\;\pi_1$\quad and \quad$\pi_2\;$. Notwithstanding 
this enlarged dependence, it may still be diagonalized by new fields 
\quad$\phi_1\;,\;\phi_2\;,\;\phi_3$\quad and \quad$\zeta\;$, the (expected) 
surprise being  
that \quad$\zeta$\quad does not couple to the source \quad$j\;$. It is a 
spurious field,
which moreover vanishes when the constraint is implemented. We skip here
the details of this derivation. 
\vfill
\eject

\sectio{\bf Conclusions}

We have shown the physical equivalence between relativistic 
HD theories of a scalar field and a reduced 2nd 
differential--order counterpart. The free part of the HD scalar theories can 
always be brought to the form (2.19) integrating by parts, and the only 
limitation of our procedure is the non-degeneracy of the resulting 
Klein--Gordon masses, i.e. we consider regular Lagrangians. The existence of a 
lower--derivative version is already suggested by the algebraic decomposition 
of the HD propagator into a sum of second--order pieces showing (physical and 
ghost) particle poles. The order--reducing program we have developed relies 
on an extension of the Legendre transformation procedure, on the use of the 
modified action principle (Helmholtz Lagrangian) and on a suitable 
diagonalization. A basic ingredient of this program is the Ostrogradski 
formalism, which we have extended to field systems.

Two approaches have been worked out. The first one follows Ostrogradski more 
closely by defining generalized momenta and Hamiltonians with a standard 
mechanical meaning, at the price of treating time separately and loosing the 
explicit Lorentz invariance. It validates a second and more powerful one 
which is explicitly Lorentz invariant. The rigorous non-invariant phase-space 
analysis strongly backs also the formal covariant methods used in HD gravity, 
where \quad$R_{\mu\nu}[g,\partial g,\partial^2g]$\quad and 
\quad$\square\, h_{\mu\nu}$\quad (in the linearized theory) are used in the 
Legendre transformation. 

The HD theories of a scalar field we have considered are generalized Klein--
Gordon theories, and hence of \quad$2N$\quad differential order according to 
the number \quad$N$\quad of KG operators involved. While the non--invarint 
approach treats all the theories on the same footing, the odd \quad$N$\quad 
and the even \quad$N$\quad cases feature qualitative differences in the 
invariant method. Also the ratio of physical versus ghost fields varies. For 
even \quad$N$\quad one finds \quad$N/2$\quad fields of each type. For odd
\quad$N$\quad one has \quad$(N-1)/2$\quad ghost (physical) and 
\quad$(N+1)/2$\quad physical (ghost) fields according to the overall negative 
(positive) sign of the free part of the HD Lagrangian. The squared masses may 
be shifted by an arbitrary common ammount, since only their differences are 
involved in the procedure. Then any of them may be zero (only one in this 
case), or tachyonic. 

On the other hand, the non--invariant procedure gets exceedingly cumbersome 
already for \quad$N=3\;$, in contrast with the (more compact) invariant one 
which remains tractable up to \quad$N=4$\quad at least. Both approaches are 
applicable to higher \quad$N$\quad, only at the prize of increasing the 
length of the calculations (namely analitically diagonalizing \quad$N\times 
N$\quad matrices). An intriguing feature of the odd \quad$N$\quad cases when 
treated with the invariant method is the occurrence of a constraint on an 
otherwise overabundant set of Ostrogradski--like coordinates and momenta, 
together with a less conventional way of working out the highest field 
derivative. Ignoring the constraint causes the appearance of a spurious 
decoupled scalar field.
        
\bigskip
\bigskip

{\bf Acknowledgements}

We are indebt to Dr. J. Le\'on for the careful reading of the manuscript
and useful suggestions.

\vfill
\eject

{\bf Appendix}

The problem of finding a matrix \quad${\cal X}$\quad with the properties 
(3.21) and (3.22) can be brought to the one of diagonalizing a symmetric 
\quad$4\times 4$\quad matrix with pure real and imaginary elements. The 
procedure is somehow tricky since there is no similarity--like transformation 
that brings the symplectic matrix \quad$\Sigma$\quad to the identity matrix, 
thus preventing a plain use of the weaponry of orthonormal transformations. 
We introduce the diagonal matrices
\quad$f\equiv diag(i,1,1,-i)$\quad and \quad$g\equiv diag(1,i,i,-1)$\quad so 
that 

$$\Sigma = gKf \quad,\, {\rm where}\quad
   K\equiv\left(\matrix{0&1&0&0\cr
                        1&0&0&0\cr
                        0&0&0&1\cr
                        0&0&1&0\cr}\right)\quad.$$\hfill (A.1)

\noindent{Taking} \quad$f\neq g$\quad does not compromise the uniqueness of 
the transformation \quad$\Phi\rightarrow\Theta$\quad as shown at the end.

Now we transform the symmetric matrix \quad$K$\quad into the \quad$4\times 
4$\quad identity by a similarity transformation

$${\cal D}={1\over{\sqrt 2}}\left(\matrix{1&1&0&0\cr
                                          -i&i&0&0\cr
                                          0&0&1&1\cr
                                          0&0&-i&i\cr}\right)
                                                 \quad ,$$\hfill (A.2)

\noindent{so} that

$${\cal D}\,{\cal K}\,{\cal D}^T 
                       ={\cal D}\,g^{-1}\Sigma\,f^{-1}\,{\cal D}^T
                       = I\quad. $$                 \hfill (A.3) 

\noindent{This} same transformation converts \quad${\cal M}_4$\quad and 
\quad${\cal M}_2$\quad into

$$\eqalign
     {{\hat{\cal M}}_4&={\cal D}\,g^{-1}{\cal M}_4 \,f^{-1}\,{\cal D}^T\cr
      {\hat{\cal M}}_2&={\cal D}\,g^{-1}{\cal M}_2 \,f^{-1}\,{\cal D}^T                                                                                           
\quad.\cr}$$ \hfill (A.4)

\noindent{Notice} that \quad${\hat{\cal M}}_2$\quad and \quad${\hat{\cal 
M}}_4$\quad are symmetric as well. This is a consequence of the vanishing of 
some critical elements in both matrices. One then verifies that they have the 
same eigenvalues, namely \quad$-i\mu M_1\;,\;i\mu M_1\;,\;i\mu M_2$\quad and
\quad$-i\mu M_2\;$, so that there exist orthogonal matrices \quad$R$\quad and 
\quad$T$\quad such that
\eject

$$T^T{\hat{\cal M}}_4T = R^T{\hat{\cal M}}_2R 
                        =i\mu\, diag(-M_1,M_1,M_2,-M_2)\;,$$  \hfill (A.5)

\noindent{while} conserving the euclidean metric \quad$I\;$:

      $$R^T I\,R = T^T I\,T =  I$$                            \hfill (A.6)

\noindent{With} the orthonormal eigenvectors as columns one obtains

$$R={1\over{2\sqrt\mu}}\left(\matrix{-R^+_1&-iR^-_1&0&0\cr
                                     -iR^-_1&R^+_1&0&0 \cr
                                       0&0&R^+_2&iR^-_2\cr 
                                      0&0&iR^-_2&-R^+_2\cr}\right)$$
                                                               \hfill (A.7)

\noindent{where} 
                
$$R^{\pm}_i\equiv{{M_i\pm\mu}\over{\sqrt{M_i}}}\quad,$$        \hfill (A.8)

\noindent{and}

$$T={1\over{2\langle 12\rangle\sqrt\mu}}
                       \left(\matrix{T^+_1&-iT^-_1&-T^-_2&iT^+_2\cr
                                     iT^-_1&T^+_1&-iT^+_2&-T^-_2\cr
                                       P^-_1&iP^+_1&P^+_2&iP^-_2\cr 
                                     iP^+_1&-P^-_1&iP^-_2&-P^+_2\cr}\right)$$
                                                                \hfill (A.9)                                                  

\noindent{where}
 
$$\eqalign{T^{\pm}_i
               &\equiv\sqrt{M_i}(\mu\sqrt{M_i}\pm\langle 12\rangle)\cr
               P^{\pm}_i
               &\equiv{{\langle 12\rangle\sqrt{M_i}}\over{M^2_i}}
                ({P\over{\langle 12\rangle}}\pm\mu M_i)\quad.\cr}$$
                                                                \hfill (A.10)                                            

\noindent{Notice} that one has pure real and imaginary matrix elements and 
vector components, and that the norm of a vector, defined as 
\quad$\mid\!V\!\mid\equiv V^TV\;$, may be imaginary as well. Since 
\quad$M^2_i\equiv m^2_i-\triangle\;$, a regularization (the dimensional one, 
for instance) is understood such that \quad$R$\quad and \quad$T$\quad have 
well defined elements.
\eject

Finally, from (A.4) and (A.5) one gets that

                         $$Y\,{\cal M}_4W={\cal M}_2\quad,$$   \hfill (A.11)

\noindent{where} \quad$W\equiv f^{-1}{\cal D}^TT\,R^T{{\cal D}^{-1}}^Tf$\quad 
and \quad$Y\equiv g\,{\cal D}^{-1}R\,T^T{\cal D}\,g^{-1}\;$. The matrix 
\quad$W$\quad has some imaginary elements and the fourth row is not 
\quad$(\,0\,1\,0\,1\,)\;$, so that it is not suitable to relate the real 
vectors \quad$\Phi$\quad and \quad$\Theta$\quad as in (3.19) yet. Moreover, 
\quad$Y\neq W^T\;$. However one may check that

$$ \left(\matrix{i&0&0&0\cr
                 0&i&0&0\cr
                 0&0&1&0\cr
                 0&0&0&1\cr}\right)\,Y = {\cal X}^T\;,\;{\rm where}\quad
  {\cal X}\equiv \;W\;\left(\matrix{-i&0&0&0\cr
                                    0&-i&0&0\cr
                                     0&0&1&0\cr
                                     0&0&0&1\cr}\right) $$      \hfill (A.12)

\noindent{is} the matrix given in (3.20), so that (A.11) writes

       $${\cal X}^T{\cal M}_4{\cal X}={\cal M}_2\quad. $$      \hfill (A.13) 

\noindent{Furthermore}, from (A.3) and (A.6) one has that

            $${\cal X}^T\Sigma\,{\cal X}=\Sigma\quad. $$      \hfill  (A.14)

\noindent{The} fourth row of \quad${\cal X}$\quad has the desired elements 
\quad$(\,0\,1\,0\,1\,)$\quad only if suitable signs are chosen for the 
eigenvectors that build up \quad$R$\quad and \quad$T\;$, so that the 
handedness of the frame is conserved by \quad${\cal X}\;$. We stress that
\quad${\cal X}$\quad is also well--defined as a differential operator, and 
that the regularization is needed only for defining the intermediate 
operators \quad$T$\quad and \quad$R^T\;$. At the end of the process the 
regularization can be put off.

\vfill
\eject

\centerline{REFERENCES}\vskip1.0cm

\noindent [1] B.Podolski and P.Schwed, {\it Rev.Mod.Phys.}{\bf 20}(1948)40.

\noindent [2] K.S.Stelle, {\it Gen.Rel.Grav.}{\bf 9}(1978)353.

             A.Bartoli and J.Julve, {\it Nucl.Phys.}{\bf B425}(1994)277.

\noindent [3] D.G.Barci,C.G.Bollini and M.C.Rocca, {\it Int.J.Mod.Phys.}
                                                      {\bf A10}(1995)1737.

\noindent [4] B.M.Pimentel and R.G.Teixeira, Preprint hep-th/9704088. 

\noindent [5] D.J.Gross and E.Witten, {\it Nucl.Phys.}{\bf B277}(1986)1. 

          R.R.Metsaev and A.A.Tseytlin, {\it Phys.Lett.}{\bf B185}(1987)52 .
 
           M.C.Bento and O.Bertolami, {\it Phys.Lett.}{\bf B228}(1989)348.

\noindent [6] N.D.Birrell and P.C.W.Davies, {\it Quantum Fields in
                 Curved Space},
           
                 Cambridge Univ.Press(1982).

\noindent [7] K.S.Stelle, {\it Phys.Rev.}{\bf D16}(1977)953.

\noindent [8] T.Goldman, J.P\'erez-Mercader, F.Cooper and M.M.Nieto,
                                    {\it Phys.Lett.}{\bf 281}(1992)219.      
            
              I.L.Buchbinder, S.D.Odintsov and I.L.Shapiro, 

                               {\it Effective Action in Quantum Gravity},                                                                         (IOP, Bristol and Philadelphia, 1992).

               E.Elizalde, S.D.Odintsov and A.Romeo,
                                     {\it Phys.Rev.}{\bf D51}(1995)1680.

\noindent [9] M.Ferraris and J.Kijowski, {\it Gen.Rel.Grav.}
                                                  {\bf 14}(1982)165.

              A.Jakubiec and J.Kijowski, {\it Phys.Rev.}{\bf D37}(1988)1406.
                                                            
              G.Magnano, M.Ferraris and M.Francaviglia,
                                  {\it Gen.Rel.Grav.}{\bf 19}(1987)465;
                                  
                                  {\it J.Math.Phys.}{\bf 31}(1990)378;
                                  {\it Class.Quantum.Grav.}{\bf 7}(1990)557.

\noindent [10] J.C.Alonso, F.Barbero, J.Julve and A.Tiemblo,
                              {\it Class.Quantum Grav.}{\bf 11}(1994)865.

\noindent [11] J.C.Alonso and J.Julve, {\it Particle contents of
                higher order gravity}, 
              
              in Classical and Quantum Gravity (proc. of the
              1st. Iberian Meeting on Gravity, 

              Evora, Portugal,1992), World Sci.Pub.Co.(1993)301.

\noindent [12] N.H.Barth and S.M.Christensen,
                                  {\it Phys.Rev.}{\bf D28}(1983)1876.

\noindent [13] K.Jansen, J.Kuti and Ch.Liu,
                                 {\it Phys.Lett.}{\bf B309}(1993)119; 
                                 {\it Phys.Lett.}{\bf B309}(1993)127.

\noindent [14] E.T.Whittaker, {\it A treatise on the analytical dynamics of
               particles and rigid bodies}, 

               Cambridge Univ. Press,(1904). Pioneeering work by 
               M.Ostrogradski and W.F.Donkin 

               is quoted here.

\noindent [15] D.J.Saunders and M.Crampin,
                                   {\it J.Phys.}{\bf A23}(1990)3169.

               M.J.Gotay, {\it Mechanics, Analysis and Geometry:
                           200 Years after Lagrange},
                      
               M.Francaviglia Ed., Elsevier Sci.Pub.(1991).

\vfill
\bye